\pdfoutput=1

\documentclass[10pt]{article}

\usepackage{listings}    
\usepackage{graphicx}    
\usepackage{subcaption}  
\usepackage{algorithm}   
\usepackage{algorithmic} 
\usepackage{booktabs}    
\usepackage{hyperref}    
\usepackage{xurl}        
\usepackage[htt]{hyphenat} 
\usepackage{microtype} 
\DisableLigatures{}    
\usepackage{adjustbox} 
\usepackage{colortbl}  

\usepackage{caption}
\usepackage{setspace}
\usepackage{multirow}
\usepackage{enumerate}
\usepackage{pdflscape}
\usepackage{pgfplots} 

\definecolor{codegreen}{rgb}{0.25,0.5,0.35}
\definecolor{codegray}{rgb}{0.5,0.5,0.5}
\definecolor{codepurple}{rgb}{0.6,0,0}
\definecolor{backcolour}{rgb}{0.95,0.95,0.92}
\definecolor{colorstring}{rgb}{0.5,0,0.35}
\definecolor{rltred}{rgb}{0.5,0,0}
\definecolor{rltgreen}{rgb}{0,0.5,0}
\definecolor{rltblue}{rgb}{0,0,0.5}
\definecolor{DarkGreen}{rgb}{0.00,0.60,0.00}
\definecolor{ScarletRed}{rgb}{0.80,0.00,0.00}
\definecolor{blizzardblue}{rgb}{0.67, 0.9, 0.93}
\definecolor{green-yellow}{rgb}{0.68, 1.0, 0.18}
\definecolor{dkgreen}{rgb}{0,0.6,0}
\definecolor{gray}{rgb}{0.5,0.5,0.5}
\definecolor{mauve}{rgb}{0.58,0,0.82}
\definecolor{lightgrey}{rgb}{0.90,0.90,0.90}
\definecolor{grey}{gray}{0.75}
\definecolor{light-gray}{gray}{0.80}

\lstdefinestyle{mystyle}{
    escapechar=©, 
	backgroundcolor=\color{backcolour},
    basicstyle=\scriptsize\ttfamily,
   	identifierstyle=\footnotesize\ttfamily,
	commentstyle=\color{codegreen},
	keywordstyle=\color{colorstring}\bfseries,
	numberstyle=\ttfamily\color{codegray},
	stringstyle=\ttfamily\color{DarkGreen},
	breakatwhitespace=false,
	breaklines=true,
	captionpos=b,
	keepspaces=true,
	numbers=left, 
	numbersep=2pt,
	showspaces=false,
	showstringspaces=false,
	showtabs=false,
	tabsize=2
}
\usepackage{xspace}
\newcommand{\evo}{{\sc EvoMaster}\xspace}
\newcommand{\etal}{{\emph{et al.}}\xspace}

\usepackage{boxedminipage}
\newenvironment{result}%
{\smallskip
	\noindent
	\let\emph=\textbf
	\begin{boxedminipage}{\columnwidth}\begin{center}\em}%
		{\end{center}\end{boxedminipage}%
	\medskip
}

\usepackage{ifthen}
\newboolean{showcomments}
\setboolean{showcomments}{true} 

\ifthenelse{\boolean{showcomments}}{
	\newcommand{\nbc}[3]{
		{\colorbox{#3}{\bfseries\sffamily\scriptsize\textcolor{white}{#1}}}
		{\textcolor{#3}{\sf\small$\langle$\textit{#2}$\rangle$}}}
	
}{
	\newcommand{\nbc}[3]{}

}

\newcommand{\totoracles}{9\xspace}
\newcommand{\foundfaults}{166\xspace}

\newcommand{\nonworkingdelete}{Non-Working Delete\xspace}
\newcommand{\sideeffectsfailedmodification}{Side-Effects Failed Modification\xspace}
\newcommand{\repeatedcreateput}{Repeated Create PUT\xspace}
\newcommand{\misleadingcreateput}{Misleading Create PUT\xspace}
\newcommand{\partialupdateput}{Partial Update PUT\xspace}
\newcommand{\nonidempotentput}{Non-Idempotent PUT\xspace}
\newcommand{\invalidmergepatch}{Invalid Merge-Patch\xspace}
\newcommand{\invalidlocation}{Invalid Location\xspace}
\newcommand{\invalidallow}{Invalid Allow\xspace}

\usepackage{authblk}                
\usepackage[square,numbers]{natbib} 
\usepackage{amsmath}                
\usepackage{amssymb}                

\usepackage{geometry}
\title{
Validating HTTP Semantics in REST APIs With Constructed Call Sequence Scenarios
}

\author[1,2]{Omur Sahin}
\author[2,3]{Andrea Arcuri}

\affil[1]{Erciyes University, Türkiye}
\affil[2]{Kristiania University of Applied Sciences, Norway}
\affil[3]{Oslo Metropolitan University, Norway}

\date{}

\begin{document}

\maketitle

\begin{abstract}
\noindent \emph{Context}:
REST APIs are widely used in industry.
These APIs use HTTP for their communications.
Failures in following the specifications of HTTP can lead to confusing and hard to use APIs,
with possibly serious software faults with dire consequences.

\noindent \emph{Objectives}:
Define novel automated techniques to automatically find HTTP semantics-level faults in existing REST APIs.

\noindent \emph{Methods}:
We extended the state-of-the-art fuzzer \evo with \totoracles new oracles to detect HTTP semantics-level faults.
Once the standard fuzzing process is finished generating $N$ test cases, a new phase is executed in which these $N$ tests
are used as a starting point to create new scenarios (i.e., new sequences of HTTP calls) aimed at validating specific HTTP properties defined in these \totoracles oracles.

\noindent \emph{Results}:
Experiments on  \totoracles artificial APIs with inject faults show that our novel techniques can successfully detect all of them.
Further experiments on 36 APIs from the WFD corpus show that our novel techniques can automatically find \foundfaults existing faults in these real-world APIs.

\noindent \emph{Conclusion}:
REST APIs use HTTP, and, as such, they need to follow its semantics to avoid misleading their clients and introducing subtle software faults.
The novel techniques presented in this paper are shown to be effective at automatically finding several of this type of faults.

\end{abstract}


\section{Introduction}

Because REST APIs are widely used in industry, lot of work has been done in academia to define techniques to automatically test this kind of APIs~\cite{golmohammadi2023testing}.
Automated test generation might aim at generating test suites that maximize different kinds of criteria, like code coverage~\cite{arcuri2019restful}, schema coverage~\cite{martin2019test} and fault detection~\cite{marculescu2022faults}.

In fuzzing, typically ``program crashes'' are used as \emph{automated oracle} for fault detection.
In the context of REST APIs, a HTTP server would not crash, even if there are faults in the business logic of the API that lead to throwing exceptions.
In those cases, the HTTP server would return a 500 (Server Error) HTTP status code.
Checking for those 500 responses is what usually used in the literature when evaluating and comparing the fault detection capabilities of REST API fuzzers~\cite{golmohammadi2023testing,Kim2022Rest,zhang2023open}.

However, there are several types of faults that do not lead to any crash.
For this reason, different types of automated oracles have been proposed and studied in the literature of fuzzing REST APIs.
These include for example \emph{robustness testing}~\cite{laranjeiro2021black}, where invalid data (based on the API schema) is sent on purpose, and the fuzzer then checks if the response of the API is correctly handled as an user-error (and not accepted as valid input).
Furthermore, many different kinds of \emph{security properties} can be evaluated during REST API fuzzing~\cite{deng2023nautilus,du2024vulnerability,arcuri2025fuzzing,sahin2026enhancing}.
The more automated oracles a fuzzer can employ, the more faults can be expected to be found when fuzzing a REST API.

REST is just a set of architectural guidelines on how to define HTTP endpoints.
Still, as REST is based on HTTP, its semantics need to be satisfied.
A REST API that does not follow HTTP would be hard to use for its clients.
For example, returning a 404 (Not Found) status code, instead of a 201 (Created), when creating a new resource with a \texttt{POST}  would be highly misleading.

Several HTTP ``rules'' can be verified already at the static level on the schema of the API~\cite{decrop2026analyzing}, e.g., based on the use of the HTTP status codes.
However, quite a few rules cannot be verified statically, as they require to make specific sequences of HTTP calls towards the API to analyze how resources are modified.

In this paper, we have analyzed the specifications of HTTP and defined \totoracles oracles to dynamically evaluate the compliance of the tested APIs with HTTP.
Our approach is as following:
given a starting set of $N$ test cases, we use those to deterministically create new tests with specific sequences aimed at evaluating those \totoracles rules;
if any rule is violated, then a HTTP-semantics fault has been found.

Technically, where those $N$ test cases come from does not matter for our approach.
Those could be existing manually written test cases, or generated on-the-fly with a fuzzer.
Still, the ``quality'' of these tests is of paramount importance for the success of our approach.
For example, if a rule requires to evaluate the effects of a successful \texttt{PUT} request, we need to have at least one test case that returns a 2xx for it.
This is not necessarily trivial, as input might have complex constraints.
Fuzz testing REST APIs to make sure to create successful 2xx calls on each endpoint is still an open research problem~\cite{zhang2023open,sahin2025wfc,arcuri2026fuzzing}.

To evaluate the effectiveness of our oracles, we extended the state-of-the-art, open-source fuzzer \evo~\cite{arcuri2018evomaster,arcuri2025tool}.
When its search process is finished (e.g., when running it for one hour), and a minimized test suite of size $N$ is created, we apply a post-processing phase in which new test cases are created and evaluated based on our HTTP-rule scenarios for the \totoracles oracles.
If any of these new tests find any HTTP-semantics fault, those new tests are added to the final output test suite given to the user at the end of the whole fuzzing process.
Note: for our implementation of these HTTP rules we used \evo because it is our own tool, but any other state-of-the-art fuzzer could had been used instead.

Our novel techniques are evaluated with two distinct sets of experiments.
First, as a sanity check, we created \totoracles artificial APIs with injected faults, for each different HTTP oracles we designed.
This set of experiments is aimed at making sure that our techniques can find faults when those faults actually exist for sure in the APIs.

Second, we aim at verifying how those oracles can find faults in real-world APIs.
To achieve that, for our experiments we use the popular corpus WFD~\cite{sahin2025wfc} (previously known as EMB~\cite{icst2023emb}).
Its latest version 4.3.0~\cite{zenodo430wfd} used in this study contains 36 distinct REST APIs, with various size and complexity.
No injected faults were added to any of these APIs.
What found in our experiments are actual faults that were already present in these APIs.

Our experiments show that it was possible to detect the injected faults in all of the \totoracles artificial APIs.
Furthermore, \foundfaults new faults were found in the WFD corpus, some of them critical.

The rest of the paper is organized as follows.
Section~\ref{sec:relatedwork} describes related work.
Our novel oracles are presented in Section~\ref{sec:httprules}.
Their evaluation in an empirical study follows in Section~\ref{sec:study}.
Threats to validity are discussed in Section~\ref{sec:threats}.
Finally, Section~\ref{sec:conclusions} concludes the paper.

%

\section{Related Work}
\label{sec:relatedwork}

There is a large body of research on the fuzzing of REST APIs~\cite{golmohammadi2023testing}.
This is an active field of research,
where many fuzzers have been proposed in the literature, like for example:
APIRL~\cite{foley2025apirl},
APIF~\cite{wang2024beyond},
ARAT-RL~\cite{kim2023adaptive},
AutoRestTest~\cite{kim2025autoresttest},
bBOXRT~\cite{laranjeiro2021black},
\evo~\cite{arcuri2025tool},
IcePick~\cite{ribeiro2026systematic},
LlamaRestTest~\cite{kim2025llamaresttest},
MINER~\cite{lyu2023miner},
Morest~\cite{liu2022icse},
Nautilus~\cite{deng2023nautilus},
ResTest~\cite{martinLopez2021Restest},
RestCT~\cite{wu2022icse},
RESTler~\cite{restlerICSE2019},
RestTestGen~\cite{viglianisi2020resttestgen},
Sche\-ma\-thesis~\cite{hatfield2022deriving},
VoAPI2~\cite{du2024vulnerability}
and
WuppieFuzz~\cite{rooijakkers2025wuppiefuzz}.

Most of these fuzzers detect faults based on 500 HTTP status code (Server Error).
Some can also detect \emph{robustness} faults (e.g., bBOXRT), whereas others can also detect \emph{security} related faults (e.g., \evo, Nautilus and VoAPI2).
Techniques have also been developed to extract likely invariants from REST API executions, and use those as automated oracles~\cite{alonso2025test}.
Other techniques rely on LLMs to exploit ``common-sense'' to define automated assertions on response fields, based on their likely natural language semantics~\cite{zhou2026restor}.

When it comes to verify HTTP compliance, there are  tools that can analyze the OpenAPI schemas of the REST APIs and flag some types of HTTP misuse.
These include tools such as SCOAS~\cite{decrop2026analyzing}, which validates 24 rules based on the use of HTTP status codes in the OpenAPI schemas.
For example, if an endpoint is declared to return a required body payload, then the status code 204 (No Content) should not be among the declared possible returned status codes.
Likewise, a \texttt{GET} endpoint should not use the 201 (Created) status code.

A web API that is compliant with HTTP is not necessarily a RESTful one.
Besides HTTP properties, also general REST design compliance can be statically evaluated, e.g., with tools such as RESTRuler~\cite{bogner2024restruler}.

Static tools that work at the OpenAPI schema level can be useful to detect several kinds of HTTP-compliance and REST design faults.
Still, an API might have faults in their implementation that are not visible at the schema level.
To detect those, \emph{dynamic} analysis in which actual HTTP calls are made towards the API is necessary.

To the best of our knowledge, there are two pieces of work that are most related to what presented in this paper.
First, RESTler defined 4 rules~\cite{atlidakis2020checking} to detect faults,
including \emph{Use-after-free} and \emph{Resource-leak rule}.
Those two rules are similar to two  presented in this paper
(described in Section~\ref{sub:nonworkingdelete} and Section~\ref{sub:sideeffectsfailedmodification}).
However, for these two rules there are several edge cases not discussed in~\cite{atlidakis2020checking}, and that we handle in this work (e.g., regarding flaky fields).
For these two rules (out of \totoracles we present in this paper), our work can be considered as an extension/improvement upon~\cite{atlidakis2020checking}.

Ribeiro \etal~\cite{ribeiro2026systematic} defined a contract specification language called Glacier to define properties on OpenAPI schemas.
They then defined a set of rules to map some HTTP-compliance checks into Glacier, and then use the tool IcePick to dynamically detect HTTP-compliance faults in REST APIs.
However, such technique has several practical limitations, as it requires the APIs to fully follow REST API design rules to be applicable.
Due to these major limitations, only two APIs from the EMB/WFD corpus could be used in the experiments, out of the 36 available at that time.
Furthermore, even in those cases, their OpenAPI schemas had to be manually modified to be able to apply IcePick on them, as discussed in~\cite{ribeiro2026systematic}.
In contrast, our techniques are more general, as they do not require any manual modification to the OpenAPI schemas, and can be successfully applied out-of-the-box on \emph{all} the 36 APIs of EMB/WFD (as we will show in Section~\ref{sec:study}).

\section{HTTP Rules}
\label{sec:httprules}

\begin{table}[!t]
  \centering
    \small
    \caption{
    Summary of the \totoracles automated oracles presented in this paper.
    We provide their temporary code in WFC format, their name, the RFC they are based on,
    and links to this article sections where these oracles are described in details.
    }
    \label{tab:rules}
        \begin{tabular}{rllc}\\
         \toprule
         Code & Name &  RFC  & Section  \\
         \midrule
         900 & \nonworkingdelete              & 9110 &  \ref{sub:nonworkingdelete}\\
         901 & \sideeffectsfailedmodification & 9110 & \ref{sub:sideeffectsfailedmodification} \\
         902 & \repeatedcreateput             & 9110 & \ref{sub:repeatedcreateput} \\
         903 & \misleadingcreateput           & 9110 & \ref{sub:misleadingcreateput} \\
         904 & \partialupdateput              & 9110 & \ref{sub:partialupdateput} \\
         905 & \nonidempotentput              & 9110 & \ref{sub:nonidempotentput} \\
         906 & \invalidmergepatch             & 5789,7386& \ref{sub:invalidmergepatch} \\
         907 & \invalidlocation               & 9110 & \ref{sub:invalidlocation} \\
         908 & \invalidallow                  & 9110 & \ref{sub:invalidallow} \\
         \bottomrule
        \end{tabular}
\end{table}

In this paper, we define  \totoracles  HTTP-compliance rules based on the execution of specific sequences of HTTP calls.
Given an existing set of $N$ tests (e.g., generated via fuzzing with tools such as \evo~\cite{arcuri2018evomaster,arcuri2025tool}, or already existing manually written tests),
our approach aims to create new test sequences, to validate those rules, starting from what available in those $N$ tests as building blocks.

Table~\ref{tab:rules} shows a summary of these rules.
Each of them is then explained in more details in the following sections.
Source code for all these rules is available in the GitHub repository of \evo.\footnote{\url{https://github.com/WebFuzzing/EvoMaster}}

To simplify the presentation of data from our empirical study, each oracle has its own unique identifying code.
We use the format from Web Fuzzing Commons (WFC)~\cite{sahin2025wfc}, where the range 9xx is reserved for work-in-progress codes that are not part yet of WFC.
In Table~\ref{tab:rules}, we also specify the source RFC those oracles are based on.
These are
``HTTP Semantics''\footnote{\url{https://datatracker.ietf.org/doc/html/rfc9110}} (RFC9110),
``PATCH Method for HTTP''\footnote{\url{https://datatracker.ietf.org/doc/html/rfc5789}} (RFC5789),
and
``JSON Merge Patch''\footnote{\url{https://datatracker.ietf.org/doc/html/rfc7386}} (RFC7386).

In the descriptions of these oracles in the following sections, there are three  common operations that are needed and are re-used several times.
These are \emph{slice}, \emph{bindAccess} and \emph{bindQueries}.
To avoid repetitions, we will define them only once, here.

The operation \emph{slice} takes as input a test case $t$ and a target call $\alpha$ in it, and removes all HTTP calls after the target $\alpha$.
For example, a test case $t$ could be composed of a sequence of five calls \texttt{POST-GET-PUT-PUT-DELETE}.
If our target $\alpha$ is the \texttt{GET}, then the result of the \emph{slice} is  \texttt{POST-GET}.

The choice of a the target $\alpha$ can be based on several different properties, like path, verb and/or returned status code.
When we need a call $\alpha$ on an endpoint that returns a specific status code, we can search for it if available in any existing test $t$ in $N$.
If so, we do not need any call \emph{after} the target $\alpha$, as those calls have no impact on the results of $\alpha$.
When we need to construct a new test that has  $\alpha$, we can hence make a copy $c$ of $t$, and \emph{slice} (i.e., remove) any call after $\alpha$.

Shorter tests are usually better, as they are easier to understand (e.g., for debugging) and take less time to execute.
For this reason, if a valid $\alpha$ is present in more than one test case in $N$, for $t$ we choose the shortest.
Removing all calls after $\alpha$ is done for the same reason, i.e., to try to have test cases as short as possible, while still satisfying any testing target we need.

Removing calls \emph{before} the target $\alpha$ would be likely wrong, as such calls might set up the state in the API that leads $\alpha$ to return what it does.
For example, a call \texttt{GET:/items/42} returning 200 might change into a 404 if its previous \texttt{PUT:/items/42} is removed.
It could feel safe to remove all other previous \texttt{GET} operations, as in ``theory'' those should not change the state of the API.
However, as the final test suite $N$ generated by \evo is already minimized, we do not do this further optimization, as it is not guaranteed to be 100\% safe (e.g., we have experienced cases of APIs wrongly 201 creating resources with \texttt{GET} operations).

When given a test case $c$ we need to add new HTTP calls in, we need to make sure that the new operations work on the same resources manipulated by the test $c$.
This is enforced with the \emph{bindAccess} operation.
For example, assume the sequence $c$ for target $\alpha$ \texttt{PUT:/items/\{id\}} returning 204:

\begin{lstlisting}
 POST   /items       -> 201
 PUT    /items/{id}  -> 204
\end{lstlisting}

Assume we want to add a new \texttt{GET} on the same resource manipulated by $\alpha$.
In this particular case, the resource is dynamically created by the \texttt{POST}.
The id selected by the server might be returned in the body payload of the response, or as part of a \texttt{Location} header.
The fuzzer would need to automatically extract such info in the generated test cases to be able to re-create the right URL paths on-the-fly when the test cases are executed.
Also, there might be more complex scenarios that could be handled recursively, like nested resources
\texttt{/a/\{id-a\}/b/\{id-b\}/c/\{id-c\}}, each one needed to be created one at a time in a chain with a different \texttt{POST} request.
Whatever strategy it is employed by the fuzzer to ``bind'' the path resolution of $\alpha$ to the resources dynamically created in the test, we use exactly the same approach when we \emph{bindAccess} the new added calls to $\alpha$.
For example, when adding a new \texttt{GET} that is \emph{bindAccess} to $\alpha$ (i.e., the \texttt{PUT} call in this example), both calls are instructed to work on the same resource dynamically created by the \texttt{POST}:

\begin{lstlisting}
 POST   /items       -> 201
 PUT    /items/{id}  -> 204
 GET    /items/{id}  -> 200
\end{lstlisting}

A further step in \emph{bindAccess}  is that we should aim at avoiding cases of 401 and 403 responses, unless we explicitly aim at them.
State-of-the-art fuzzers like \evo can create tests in which different calls in a single test can be made by different logged-in users.
This is needed for testing access policy rules~\cite{sahin2026enhancing}.
As part of \emph{bindAccess}, we make sure that the new added call uses the same authentication credentials as $\alpha$.

The verb \texttt{GET} has no defined body payload.
Still, it can have query parameters.
When adding a newly created HTTP call $\beta$ that is \emph{bindAccess} to $\alpha$, we should avoid parameters that might influence how data is retrieved, possibly in different ways of how they are set in $\alpha$.

To address this problem, in \emph{bindQueries} of $\beta$ to $\alpha$ we use the following approach.
Any query parameter that is not marked as ``required'' in the schema of $\beta$ is removed.
Then, if there is any query parameter used in $\alpha$ that has the same name as a parameter in $\beta$,
in $\beta$ we copy the value from $\alpha$, if they are of the same type (e.g., we do not copy a string value into a boolean parameter, even if two variables have the same name).
For all other cases (i.e., required parameters in $\beta$ with no matching name in $\alpha$), we randomly sample valid inputs (e.g., based on the variable type, numeric range and/or regex constraints).

Note that, in the following sections, we only provide  high-level descriptions of our novel algorithms to detect these types of HTTP-compliance faults.
Several low level details (e.g., how to deal with payloads in JSON, XML and \texttt{x-www-form-urlencoded} forms format) are not discussed here.
The interested reader is referred to our open-source implementation for all these details.

\subsection{\nonworkingdelete}
\label{sub:nonworkingdelete}

A correct \texttt{DELETE} operation should remove a resource at the given provided URL path, returning a status code in the 2xx range (RFC9110).
If the resource did not exist already, a \texttt{DELETE} should return a 404 (Not Found), but it might also still return a success status code 204 (No Content).
After a resource has been deleted, a \texttt{GET} operation on it should fail (i.e., returning a status code not in 2xx, typically a 404).

To verify the correctness of the \texttt{DELETE} endpoints, we do as follows:

\begin{enumerate}

\item For each \texttt{DELETE} endpoint $d$ in the schema having as well a declared \texttt{GET} endpoint $g$ on it, find a test case $t_d$ in $N$ using such a \texttt{DELETE} returning a 2xx.
      If more than one test case is found, choose the shortest (in terms of HTTP calls).
      If none is found, then this oracle check is skipped for such endpoint.

\item From the test case $t_d$, \emph{slice} a new copy $c_d$, where calls after target $\alpha$ \texttt{DELETE} (if any) are removed.
      The last HTTP call $\alpha$ in $c_d$ would hence be a \texttt{DELETE} returning a 2xx status code when $c_d$ is executed.

\item Create a new \texttt{GET} call $\beta$ for $g$, \emph{bindAccess} and \emph{bindQueries}  it to $\alpha$.
    Such new HTTP call $\beta$ is added to $c_d$ in second last position, just before the target $\alpha$.
    If $c_d$ already had a \texttt{GET} operation satisfying all these constraints, then this step is skipped.

\item Make a copy of $\beta$ called $\gamma$, and append it to the end of $c_d$, after the last call $\alpha$. Let us call this new test case $k_d$.

\item   If when executing $k_d$ we have $\beta$ \texttt{GET}, $\alpha$ \texttt{DELETE} and  $\gamma$ \texttt{GET} returning a 2xx status code, then the $d$ endpoint is defective, and a ``\nonworkingdelete'' fault has been identified by the test case $k_d$.
        Such $k_d$ can then be added to $N$.

\end{enumerate}

For example, if $t_d$ is something like:

\begin{lstlisting}
GET    /items       -> 200
POST   /items       -> 201
DELETE /items/{id}  -> 204
PUT    /items/{id}  -> 201
\end{lstlisting}

then, the resulting $k_d$ would be:

\begin{lstlisting}
GET    /items       -> 200
POST   /items       -> 201
GET    /items/{id}  -> ?
DELETE /items/{id}  -> 204
GET    /items/{id}  -> ?
\end{lstlisting}

By having a successful \texttt{GET} before the \texttt{DELETE}, we make sure to check that the resource exists.
If it still exists after the \texttt{DELETE}, and it can be retrieved again with a \texttt{GET}, then the endpoint is faulty.

\begin{figure}
\begin{lstlisting}[language=Java,numbers=left,xleftmargin=2em]
/**
* Calls:
* 1 - (200) PUT:/api/resources/{id}
* 2 - (200) GET:/api/resources/{id}
* 3 - (204) DELETE:/api/resources/{id}
* 4 - (200) GET:/api/resources/{id}
* Found 1 potential fault of type-code 900 ©\label{line:delete-summary}©
*/
@Test @Timeout(60)
fun test_3_getOnResourcDeleteDoesNotWork()  {

    given().accept("*/*")
        .header("x-EMextraHeader123", "")
        .put("${baseUrlOfSut}/api/resources/291")
        .then()
        .statusCode(200)
        .assertThat()
        .body(isEmptyOrNullString())

    given().accept("*/*")
        .header("x-EMextraHeader123", "42")
        .get("${baseUrlOfSut}/api/resources/291?EMextraParam123=42")
        .then()
        .statusCode(200)
        .assertThat()
        .contentType("text/plain")
        .body(containsString("Data for 291"))

    // Fault900. Resource Still Accessible After Successful DELETE.  ©\label{line:delete-call}©
    given().accept("*/*")
        .header("x-EMextraHeader123", "42")
        .delete("${baseUrlOfSut}/api/resources/291?EMextraParam123=42")
        .then()
        .statusCode(204)
        .assertThat()
        .body(isEmptyOrNullString())

    given().accept("*/*")
        .header("x-EMextraHeader123", "42")
        .get("${baseUrlOfSut}/api/resources/291?EMextraParam123=42")
        .then()
        .statusCode(200)
        .assertThat()
        .contentType("text/plain")
        .body(containsString("Data for 291"))
}
\end{lstlisting}
\caption{\label{fig:900}
Example of generated test for an artificial API in which a ``\nonworkingdelete'' fault is correctly identified.
}
\end{figure}

Finding faults is important.
Enabling users to debug and finally fix the founded faults is important as well.
For this goal, in \evo we  generate fully executable test cases in different programming languages
(e.g., Java, Kotlin, Python and JavaScript).
Found faults are directly marked in the generated tests, via code comments.
Figure~\ref{fig:900} shows an example of generated test in Kotlin for an artificial API where a \texttt{DELETE} endpoint is wrongly implemented.
With our techniques, this test case reveals the presence of the fault, as highlighted on Line~\ref{line:delete-summary} and Line~\ref{line:delete-call}.

\subsection{\sideeffectsfailedmodification}
\label{sub:sideeffectsfailedmodification}

If a \texttt{PUT} or \texttt{PATCH} operation on a specific resource fails due to a user error (i.e., status code in the 4xx range), there should be no side-effect on the resource (RFC9110).
Inside a single operation, either the modifications on a resource are all applied, or none.
If an operation fails due to an invalid input element (e.g., resulting in a 400 response), then there should be no partial update for the other input elements.
Operations are expected to be atomic.

To verify that the resource has not been changed, a possibility would be to add one \texttt{GET} request before the failed \texttt{PUT}/\texttt{PATCH}, and then another repeated \texttt{GET} after it.
The results of these two \texttt{GET}s can then be compared.
If the   \texttt{PUT}/\texttt{PATCH} failed, then the two \texttt{GET}s should return the same result, otherwise the failed modification might had side-effects.
However, there are two major problems here that must be handled:
(1) how to construct the \texttt{GET}s, which depends on the reason why the \texttt{PUT}/\texttt{PATCH} failed (e.g., 401, 403 and 404 would need to be handled differently from the other 4xx cases);
and
(2) how to deal with the possible flakiness and non-determinism in the responses of the \texttt{GET}s.

In our approach, we do the following:

\begin{enumerate}

\item  Consider each path $p$ defined in the schema having a \texttt{PUT}/\texttt{PATCH} and a \texttt{GET} endpoint.

\item Check in $N$ for any test $t_z$ having an $\alpha$ \texttt{PUT}/\texttt{PATCH}  call on $p$, considering four different cases: 401, 403, 404 and any of the other 4xx. For each of those four cases, if any is found, make a copy $c_z$ for that \texttt{PUT}/\texttt{PATCH}  call $\alpha$.
    If more than one test case fits those constraints, choose the shortest.
    If none is found, then this oracle check is skipped for such endpoint.

\item If $\alpha$ returns 404, \emph{slice} in $c_z$ any call after $\alpha$.
    Create two duplicated \texttt{GET} requests on $p$, which are \emph{bindAccess} and \emph{bindQueries} for $\alpha$.
    One  \texttt{GET} is called before $\alpha$, and the other after.
    In this final updated $c_z$, the two added \texttt{GET} calls should return 404 as well, otherwise it means that $\alpha$ had side effects (e.g., creating or deleting the resource).
    The case of 404 is special compared to the other three, as there is no response content to compare.

\item In the other three cases (i.e., 401, 403 and all other 4xx),
       find in $N$ a test case $t_g$ that returns 2xx for the \texttt{GET} endpoint on $p$.
       If more than one test case fits those constraints, choose the shortest.
       If none is found, then this oracle check is skipped for such endpoint.

\item Copy $t_g$ and \emph{slice} all calls after the target \texttt{GET} for $p$, resulting in the test $c_g$.
      By returning 2xx on a \texttt{GET}, we make sure we consider a case in which the resource at $p$ exists.

\item From $t_z$, make a copy of only the single HTTP call $\alpha$, with no other calls before or after it.
     Add this copy of a single HTTP call to $c_g$, and \emph{bindAccess} to the target \texttt{GET} for $p$.

\item If $\alpha$ returned 401, remove any authentication info for it in the copy inside $c_g$ (which might had been added when doing the \emph{bindAccess}).

\item If $\alpha$ returned 403, we aim at having that it should still return a 403 when copied and added to $c_g$.
    A randomly chosen authentication credential (among the ones available given as input to the fuzzer) that is different from the one used in the \texttt{GET} is selected.

\item If $\alpha$ returned a 4xx that is not a 401, 403, or 404, then there is no need to make any further change to the authentication credentials (as \emph{bindAccess} would set the same credentials as the \texttt{GET}).

\item For the 401, 403 and 4xx cases, make a copy of 2xx \texttt{GET}, and append it to $c_g$ after the $\alpha$.
    If the content returned in these two \texttt{GET} are ``different'', then the $\alpha$ call had side-effects, and a ``\sideeffectsfailedmodification'' fault has been found.
    Such $c_g$ can then be added to $N$.

\end{enumerate}

The checking of ``differences'' between two \texttt{GET} responses on the same resolved endpoint resource might be problematic.
This is due to the fact that some fields might be a source of flakiness and/or non-determinism.
For example, if a field represents the current time, calling the \texttt{GET} twice would return different results.
To try to minimize false positives when checking if failed $\alpha$ \texttt{PUT}/\texttt{PATCH} had side-effects, we do the following.

\begin{enumerate}

\item Let $B$ be the response of the first \texttt{GET}, $M$ be the payload sent by $\alpha$, and $A$ be the response of last \texttt{GET} after the execution of $\alpha$.

\item If $A$ is missing field values that were present in $B$, then it means $\alpha$ wrongly deleted them.

\item If $A$ has any field $f$ with value that is different from $B$, i.e., $A.f\neq B.f$, then, to avoid flakiness issues, we check what sent in $\alpha$.
    If $A.f = M.f$ while being $A.f\neq B.f$, then it means that specific modification was wrongly applied although  $\alpha$ failed.

\end{enumerate}

\subsection{\repeatedcreateput}
\label{sub:repeatedcreateput}

A successful \texttt{PUT} request that creates a new resource \emph{must} return a status code of 201 (RFC9110).
If instead it updates an existing resource, it \emph{must} return either a 200 or 204 status code (RFC9110).

If in a test case a \texttt{PUT} request returns 201, then re-executing the same operation with another \texttt{PUT} operation on the same resource must not return a 201.
If it does, then it is a HTTP-semantics fault.

To verify this rule, we follow this procedure:

\begin{enumerate}

\item Consider each path $p$ with a \texttt{PUT} operation for it.

\item Find in $N$ any test case $t_p$ with a \texttt{PUT} on $p$ returning status code 201.
      If more than one test case fits those constraints, choose the shortest.
      If none is found, then this oracle check is skipped for such endpoint.

\item Create a copy $c_p$ of $t_p$, and \emph{slice} any call after the target $\alpha$ \texttt{PUT}.

\item Make a copy of $\alpha$, and append it to $c_p$.

\item If when evaluating $c_p$ the last two \texttt{PUT} calls on the same resource both return 201, then a ``\repeatedcreateput'' fault has been found.
    Such $c_p$ can then be added to $N$.

\end{enumerate}

\subsection{\misleadingcreateput}
\label{sub:misleadingcreateput}

As explained in the description of ``\repeatedcreateput'' (Section~\ref{sub:repeatedcreateput}), a successful \texttt{PUT} must either return a 201 if it creates the resource, or a 200/204 if it updates it.

An alternative way to verify this property is to rely on the use of \texttt{GET}, if any is present.
If a \texttt{GET} call shows that a resource exists (e.g., status 2xx), then a \texttt{PUT} on it should not create a new resource (i.e., not returning a 201).
Note that this oracle ``\misleadingcreateput'' can potentially find faults that ``\repeatedcreateput'' cannot, and vice-versa, based on how those endpoints are implemented.
Furthermore, failing scenarios might point to faults in other endpoints.
For example, if a \texttt{GET} wrongly returns always 200 even in case of errors, this oracle would detect a fault, although the fault would be in the \texttt{GET} endpoint and not in the \texttt{PUT}.

To verify this rule, we use the following approach:

\begin{enumerate}

\item Consider each path $p$ with both a \texttt{PUT} and \texttt{GET} operation for it.

\item Find in $N$ a test case $t_g$ where a \texttt{GET} on $p$ returns 2xx, and a test case $t_z$ where a \texttt{PUT} on $p$ returns 201.
    If more than one test case fits those constraints, choose the shortest.
    If none is found, then this oracle check is skipped for such endpoints.

\item Make a copy $c_g$ of $t_g$, and \emph{slice} all calls after $\beta$ \texttt{GET}.

\item Make a copy of the single call $\alpha$ \texttt{PUT} in $t_z$, append such copy to $c_g$, and  \emph{bindAccess} it to $\beta$ (so the two operations work on the same resource).

\item If, when executing $c_g$, its last call $\alpha$ \texttt{PUT} returns a 201, then a ``\misleadingcreateput'' fault is identified.
    Such $c_g$ can then be added to $N$.

\end{enumerate}

\subsection{\partialupdateput}
\label{sub:partialupdateput}

A successful \texttt{PUT} request must make a full replacement of the resource, with no partial updates (RFC9110).
For example, if a resource is an object with several fields, all of them should be modified.
If some fields are optional, and they are not set in the body payload of a \texttt{PUT}, that is equivalent to put them to ``null''.
Existing data for those fields should hence be deleted.

If a user wants to make some changes just for some fields in a resource, then using a \texttt{PUT} is wrong.
For that use-case, a \texttt{PATCH} with a \emph{JSON Merge Patch} (RFC7386) should be used instead.

To detect faults related to \texttt{PUT} making partial updates, we do as follow:

\begin{enumerate}

\item Consider each path $p$ with both a \texttt{PUT} and \texttt{GET} operation for it.

\item Find in $N$ a test case $t_z$ where a $\alpha$ \texttt{PUT} on $p$ returns 2xx.
      If more than one test case fits those constraints, choose the shortest.
      If none is found, then this oracle check is skipped for such endpoint.

\item Create a copy $c_z$ of $t_z$, and \emph{slice} all calls after $\alpha$.

\item Sample a new HTTP call $\beta$ \texttt{GET} for $p$, append it to $c_z$,  \emph{bindAccess} and \emph{bindQueries} it to $\alpha$.

\item Compare the field contents of the input payload of the $\alpha$ \texttt{PUT} and the response of the $\beta$ \texttt{GET} to verify if \texttt{PUT} wrongly did a partial-update.
    If so, a ``\partialupdateput'' fault is detected.
    Such $c_z$ can then be added to $N$.

\end{enumerate}

A full field-by-field comparison between these two objects would most likely lead to false positives.
The shape of the objects (i.e., which fields it contains) between what sent by the \texttt{PUT} and what returned in the \texttt{GET} might not be the same.
The \texttt{PUT} might contain sensitive fields that are not for ``reading'' (i.e., the \texttt{GET} would not return them).
Likewise, the \texttt{GET} might contain fields that are created server-side (e.g., ids and timestamps) and that are not modifiable via a \texttt{PUT}.

To avoid these potential issues in our object matching, we look at the provided OpenAPI schema, and only verify the equivalence of fields that are declared in \emph{both} endpoints.
Fields that are source of flakiness (e.g., server-side generated ids and live-timestamps) are less likely to be set programmatically by the API clients (e.g., via a \texttt{PUT} request).
This should minimize the presence of false positives (e.g., wrongly declaring a ``\partialupdateput'' fault due to a flaky field), albeit it cannot formally guarantee their total absence.

\subsection{\nonidempotentput}
\label{sub:nonidempotentput}

In HTTP, the method \texttt{PUT} is \emph{idempotent} (RFC9110):
\emph{``A request method is considered ``idempotent'' if the intended effect on the server of multiple identical requests with that method is the same as the effect for a single such request''}.
This means that executing the same \texttt{PUT} one or more times \emph{must} have the same final result on the server (albeit the responses might be different, e.g., a 201 on first call, with a 204 on each following repeated call).
Non-idempotent methods are for example \texttt{POST} and \texttt{PATCH} (RFC9110).

A \texttt{PUT} endpoint that is wrongly implemented in a non-idempotent way not only would make the API misleading and harder to use for clients, but it also can lead to severe failures in the business logic of the API.
Idempotency is something that is known and handled throughout the whole HTTP stack, from HTTP clients to routers and API gateways.
The critical aspect here is that \emph{any} entity between the client and API in a \texttt{PUT} call can arbitrarily decide to re-execute any idempotent endpoint.
This might happen for a variety of reasons, like for example connection timeouts and message-buffer issues.
A client might send a single \texttt{PUT} request, whereas the API might receive it two or more times.

For idempotent methods, the fact that the API might receive duplicates of a request is not a problem, apart from wasted computational time.
For non-idempotent methods, those  should never be automatically repeated, and rather errors should be returned in case of network problems.

Whether the semantics of an endpoint is idempotent or not depends on how the API is implemented.
For example, consider the case of a \texttt{PUT} endpoint for a path \texttt{/accounts/\{id\}/deposit}.
In this example, \emph{adding} a specified amount of currency to the user's balance via a \texttt{PUT} would be a catastrophic software fault, as such semantics is not idempotent (a \texttt{POST} should had been used instead).
On the other hand, \emph{setting} the balance via a \texttt{PUT} endpoint for a path \texttt{/accounts/\{id\}/balance} would be fine.
Setting a value to $x$ one or fifty times would still result in the same final $x$.

What makes this ``\nonidempotentput'' fault so insidious is not just its potential catastrophic consequences, but also how hard would it be to debug it, especially if a developer does not understand idempotency.
For example, the fault (in which the non-idempotent side-effect of a \texttt{PUT} is executed more than once for one single request) might appear randomly (due to physical network) only once every $K$ calls on average, where $K$ could be $10 000$.
This is not an hypothetical problem, but a very concrete one which is not uncommon (as for example we have experienced directly in the development and testing of \evo).

To detect this type of fault, we do as follows:

\begin{enumerate}

\item Consider each path $p$ with a \texttt{PUT} endpoint.

\item Check if there is any \texttt{GET} endpoint on $p$, or any of its ancestor $p'$.
      If so, choose the closest one (best if $p'=p$).
      For example,  if \texttt{/accounts/\{id\}/deposit} has no \texttt{GET} operation, but there is one for \texttt{/accounts/\{id\}}, then we use it for $p'$.

\item Find in $N$ a test case $t_z$ where a $\alpha$ \texttt{PUT} on $p$ returns 2xx.
      If more than one test case fits those constraints, choose the shortest.
      If none is found, then this oracle check is skipped for such endpoint.

\item Make a copy $c_z$ of $t_z$, and \emph{slice} any call after the $\alpha$.

\item Sample a $\beta$ \texttt{GET} call on $p'$,  append it to $c_z$,  \emph{bindAccess} and \emph{bindQueries} it to $\alpha$.

\item Duplicate the last two \texttt{PUT}-\texttt{GET} calls, and append them to $c_z$.

\item If all these calls are successful (i.e., returned 2xx), then check  the returned content of the two duplicated \texttt{GET} requests to see if the implementation of the \texttt{PUT} was not idempotent.
    In such a case, a ``\nonidempotentput'' fault is detected.
    Such $c_z$ can then be added to $N$.

\end{enumerate}

Given a starting $t_z$ like:

\begin{lstlisting}
POST   /accounts               -> 2xx
PUT    /accounts/{id}/deposit  -> 2xx
\end{lstlisting}

then the resulting $c_z$ would be:

\begin{lstlisting}
POST   /accounts               -> 2xx
PUT    /accounts/{id}/deposit  -> 2xx
GET    /accounts/{id}          ->   ?
PUT    /accounts/{id}/deposit  ->   ?
GET    /accounts/{id}          ->   ?
\end{lstlisting}

where all these operations work on the same resource dynamically generated in the first \texttt{POST} call.

The challenge here is to how to compare the results of the two \texttt{GET}s without resulting in false positives due to flaky fields.
As the affected fields in the \texttt{GET}s might not be directly related to what sent in the \texttt{PUT}s (e.g., ``balance'' vs.~``deposit''), a full field-by-field comparison is unwise.

There is a balance to strike between fault detection effectiveness and false positive minimization.
For this study, we chose the following approach:
we only compare the equivalence of numeric and boolean fields, and,
 for arrays, we check their size, but not their content.
We explicitly ignore to compare string fields, as those are usually the major source of flakiness in the results (e.g., timestamps).

If in this restricted set of comparisons, between the results of these two \texttt{GET}s, there is any difference, then a ``\nonidempotentput'' fault has been found, as this scenario would have a high chance of demonstrating that the implementation of the \texttt{PUT} is not idempotent.

\subsection{\invalidmergepatch}
\label{sub:invalidmergepatch}

A \texttt{PATCH} operation is used to define arbitrary partial updates to resources (RFC5789)
The \texttt{PATCH} standards itself does NOT define how modifications have to be carried out.
What modifications to apply is based on the content type of the body payload of the \texttt{PATCH}.
An API can define its own custom formats, or use standards such as
\emph{JSON Patch} with type \texttt{application/json-patch+json} (RFC6902),
or
\emph{JSON Merge Patch} with type \texttt{application/merge-patch+json} (RFC7386).

At a high level, a JSON Merge Patch works like a \texttt{PUT} request, with the difference that the updates are partial.
For example, if a resource is an object with several fields, a JSON Merge Patch could provide an object payload with only the fields to modify.
Fields that are not specified in the \texttt{PATCH} should be left untouched.
If those unspecified fields are modified, then the JSON Merge Patch is incorrectly implemented.
This ``\invalidmergepatch'' rule for \texttt{PATCH} could be considered as the direct opposite of ``\partialupdateput'' for \texttt{PUT} (recall Section~\ref{sub:partialupdateput}).

What makes a JSON Merge Patch potentially tricky to implement, and so potentially prone to software faults, is how \emph{null} and \emph{undefined} are treated in statically-typed languages.
In contrast to JSON,  programming languages like Java or Kotlin have no concept of \emph{undefined}.
JSON objects such as \texttt{\{\}} and \texttt{\{"x":null\}} are different in JSON, but would map to the same Data Transfer Object (DTO) instance in Java (e.g., \texttt{class DTO\{String x;\}}), with \texttt{x=null}.
The problem here is that JSON Merge Patch (RFC7386) treats \emph{null} and \emph{undefined} differently.
A field marked as \emph{null} should be deleted, whereas one marked as \emph{undefined} (i.e., left unspecified) should not be touched (RFC7386).

In programming languages such as Java, when implementing APIs supporting RFC7386, the solution is to either parse the \texttt{PATCH} payloads manually without mapping them to DTOs (which is error prone), or use advanced mechanisms like libraries using \texttt{Optional} in Java with Jackson to specially handle \emph{undefined}, like \texttt{class DTO\{Optional<String> x;\}}.
Other programming languages and JSON parsing libraries will need to have their own special ways to deal with \emph{undefined}.
If neither of these options is done, then the implementation of the JSON Merge Patch endpoints would be most likely faulty.

To verify this rule, we use the following approach:

\begin{enumerate}

\item Consider each path $p$ with a \texttt{PATCH} endpoint that uses the input type \texttt{application/merge-patch+json}.

\item If none is found, rather consider any \texttt{PATCH} that uses the undefined \texttt{application/json} (which has no formal semantics for  \texttt{PATCH}). The motivation here is that \emph{JSON Merge Patch} is the easiest, most common way to implement a \texttt{PATCH}, and developers might have simply been unaware of the need for specific type \texttt{application/merge-patch+json}.
If still none is found, then this oracle check is skipped for such path $p$.

\item Find in $N$ a test case $t_z$ where a $\alpha$ \texttt{PATCH} on $p$ returns 2xx.
      If more than one test case is found, choose the shortest (in terms of HTTP calls).
      If none is found, then this oracle check is skipped for such endpoint.

\item Create a copy $c_z$ of $t_z$, and \emph{slice} all calls after $\alpha$.

\item Sample a new HTTP call $\beta$ \texttt{GET} for $p$, add it to $c_z$ before $\alpha$,  \emph{bindAccess} and \emph{bindQueries} it to $\alpha$.

\item Make a copy $\gamma$ of the $\beta$ \texttt{GET}, and append it to $c_z$.

\item Compare the field contents of the payload of the $\alpha$ \texttt{PATCH} and the response of the two \texttt{GET}s to verify if \texttt{PATCH} wrongly modified fields that should not had been touched.
    If it happens, then  a ``\invalidmergepatch'' fault has been found.
    Such $c_z$ can then be added to $N$.

\end{enumerate}

The first call to \texttt{GET} gives the current state of the resource $X_b$ before the \texttt{PATCH}.
The second call to \texttt{GET} gives the resulting resource $X_a$ after it has been updated.
In the \texttt{PATCH}, we check which fields $F$ are not set, i.e., the fields that are part of the OpenAPI schema for that object, but are not included in the sent JSON payload (this is different from them being set to \texttt{null}).
If  for any field $f \in F$ there is a difference between the two \texttt{GET}s, i.e., $X_b.f\ne X_a.f$, then a fault is identified.
Note that we ignore differences in the fields that are in the \texttt{GET} but that cannot be modified with the \texttt{PATCH}, as those might be potential sources of flakiness.

\subsection{\invalidlocation}
\label{sub:invalidlocation}

A response to an API call might contain a HTTP \texttt{Location} header.
In HTTP, this is used for two main reasons (RFC9110):
(1) to enable automated redirection in browsers (using status code in the 3xx range),
and (2) to specify where a newly created resource (e.g., a 201 on a \texttt{POST}) can be located when ids are generated on the server-side.

Many client libraries automatically follow 3xx redirections.
If the link provided in the  \texttt{Location} header is wrong (e.g., misspelled or no longer valid), then the automated redirection would fail.
In the case of newly generated resources on a 201 response, one would have to manually make a call to the link provided in  \texttt{Location} header.
However, HTTP does not specify what operations should be available on such links (e.g., \texttt{GET} and \texttt{DELETE}).
Furthermore, the link might point to external services outside the tested API.

To verify that those links are correct, we do as follow:

\begin{enumerate}

\item Consider each endpoint $e$ defined in the schema.

\item For each endpoint $e$, find in $N$ a test case $t_l$ in which a call $\alpha$ has response containing a \texttt{Location} header $l$.
      If more than one test case is found, choose the shortest (in terms of HTTP calls).
      If none is found, then this oracle check is skipped for such endpoint $e$.

\item Make a copy of $c_l$ of $t_l$, \emph{slice} any HTTP calls after the $\alpha$ returning $l$.

\item If $l$ points to an external service, append to $c_l$ a \texttt{GET} call $\beta$ towards $l$.

\item If $l$ points towards the tested API (e.g., a relative path), match in the schema all the endpoints for that path.
      For example, if $l$ is \texttt{http://localhost:8080/items/42}, then it could match the endpoints \texttt{GET:/items/\{id\}} and \texttt{DELETE:/items/\{id\}}, but not \texttt{POST:/items}.
      Choose one matched, available endpoint, based on this verb priority:
      \texttt{GET}, \texttt{DELETE}, \texttt{POST}, \texttt{PUT}, \texttt{PATCH}.
      If in the schema there are \emph{required} elements for the selected endpoint (e.g., query parameters and body payloads), those are randomized, but still valid according to the schema (i.e., within the validity of any defined constraints, like numeric ranges and regular expressions).
      Append such new call $\beta$ to $c_l$.

\item If when executing $c_l$ the call $\beta$ to $l$ returns either a
      404 (Not Found),
      405 (Method Not Allowed),
      500 (Server Error)
      or 501 (Not Implemented),
      then a ``\invalidlocation'' fault has been identified.
      Such $c_l$ can then be added to $N$.

\end{enumerate}

\subsection{\invalidallow}
\label{sub:invalidallow}

An OpenAPI schema would define which endpoints are available in the described API.
In HTTP, given a path $p$, the method \texttt{OPTIONS} can be used to check which verbs are available on that resource (RFC9110).
In such cases, the server \emph{might} respond with a \texttt{Allow} header specifying each valid verb present on that resource.
Usually, a developer would not need to implement the method \texttt{OPTIONS} manually, but rather it would be automatically handled by the HTTP servers based on the actual available endpoints.

When an OpenAPI schema is provided, the use of \texttt{OPTIONS} might be redundant, as all needed info for the clients is already specified in the schema.
However, it might be problematic if what returned in the \texttt{OPTIONS} is not consistent with what declared in the schema.
The returned \texttt{Allow}  might miss verbs that are in schema, as well as having extra verbs not in the schema.
This latter point might also become a security vulnerability~\cite{sahin2026enhancing}: those endpoints might not be  meant for the general public (e.g., debugging, work-in-progress or deprecated endpoints), but still be wrongly accessible.

To check for these possible inconsistency errors, we do as follow:

\begin{enumerate}

\item  For each path $p$ declared in the schema, create a new test case $c_o$, consisting of a single \texttt{OPTIONS} call on $p$.

\item  If the response contains an \texttt{Allow} header (which is not mandatory in HTTP), check its content.

\item  While ignoring verbs such as \texttt{HEAD} and \texttt{OPTIONS} itself, if there is any discrepancy (i.e., missing and/or extra verbs) with what declared in the schema, then a ``\invalidallow'' fault has been found.
    Such $c_o$ can then be added to $N$.

\end{enumerate}

\section{Empirical Study}
\label{sec:study}

In this paper, we aim at answering the following research questions:

\begin{itemize}

\item {\bf RQ1}: Do our novel techniques presented in this paper reliably find injected faults in artificial example APIs?

\item {\bf RQ2}: What faults can our novel techniques find when fuzzing real-world APIs with \evo?

\item {\bf RQ3}: What is the computational overhead of our novel techniques?

\end{itemize}

\subsection{Fuzzer Integration}
\label{sub:}

All our novel oracles described in Table~\ref{tab:rules} have been implemented in the open-source fuzzer \evo~\cite{arcuri2018evomaster,arcuri2025tool}.
When \evo is used to fuzz test a REST API, at the end of its fuzzing session a new phase is executed to construct new test cases to find HTTP-compliance faults based on the \totoracles rules defined in this paper.

These test cases are based on the output $N$ tests generated by \evo during the fuzzing session.
If any new test case find new fault, those tests are added to $N$ before giving the final output to the users.
This is exactly the same approach we do when generating \emph{security} tests in \evo after the main fuzzing session is completed~\cite{sahin2026enhancing}.

A fuzzer can be run for any arbitrary amount of time, like 10 minutes, 1 hour or 24 hours.
The longer a fuzzer is left running, the higher the likelihood of getting better final results (e.g., higher code coverage and more crashes).
However, our novel techniques build new test cases deterministically based on the provided input set $N$.
The number of evaluated test cases is a constant depending on the size and properties of $N$.

Therefore, we simply execute such phase after the main fuzzing session, but provide a pre-emptive timeout if such new phase takes too long.
In our current version of \evo, such timeout is based on $X\%$ of the fuzzing budget, e.g., $10\%$.
For example, if \evo is instructed to fuzz an API for 60 minutes (i.e., 1 hour), then the follow-up, HTTP-compliance-checks phase will be run for at most 6 minutes.
However, as we will show in the empirical study, such new phase takes at  most a few seconds in most of the cases.

\subsection{Case Study}
\label{sub:}

To answer our research questions, we carried out two different sets of experiments.

For answering {\bf RQ1}, we created \totoracles distinct artificial APIs with manually injected faults, for each different automated oracles presented in this paper.
These APIs are simple, without any confounding factors such authentication, complex input constraints or requiring databases.
They are written in Kotlin, using SpringBoot.

The motivation here is that, if our novel techniques work, then they should reliably find all these faults.
Finding such faults should be done so reliably that those APIs could be added as end-to-end tests of \evo itself~\cite{arcuri2023building}.
Otherwise, if our techniques could not find those, then there would be no point in trying to analyze more complex, real-world APIs.
In other words, this first set of experiments can be considered as a viability check of these techniques.

Another important reason for using artificial examples with known injected faults is that no fault can be found if they do not exist.
On the one hand, if a novel technique does not find any fault on a real-world API, it might be simply because there is no fault to find, instead of being due to the efficiency of the proposed technique.
On the other hand, if no fault of a specific category is found in an empirical study, it could be that type of faults is not common in practice, or that the proposed technique is not good at finding them in real-world APIs.
For these reasons, experiments on \emph{both} artificial and real-world APIs are needed.

For each of these \totoracles artificial APIs, we ran \evo 5 times, up to 1 minute, and verified whether the inject faults were found.

\begin{table}[!t]
\centering
\caption{ Statistics of the employed 36 REST APIs in our empirical study,
including number of source files, numbers of lines of code (LOCs), the number of HTTP endpoints,
whether they require authentication, and the databases they use (if any).
\label{tab:suts}
}
\vspace{-1.5\baselineskip}
\resizebox{1.0\textwidth}{!}{
\begin{tabular}{l rrrrr}\\
\toprule
SUT & \#SourceFiles & \#LOCs & \#Endpoints & Auth& Databases\\
\midrule
\emph{bibliothek} &  33 &  2176 &  8 &  &  MongoDB \\
\emph{blogapi} &  89 &  4787 &  52 & \checkmark &  MySQL \\
\emph{catwatch} &  106 &  9636 &  14 &  &  H2 \\
\emph{cwa-verification} &  47 &  3955 &  5 &  &  H2 \\
\emph{erc20-rest-service} &  7 &  1378 &  13 &  &   \\
\emph{familie-ba-sak} &  1089 &  143556 &  183 & \checkmark &  PostgreSQL \\
\emph{features-service} &  39 &  2275 &  18 &  &  H2 \\
\emph{genome-nexus} &  405 &  30004 &  23 &  &  MongoDB \\
\emph{gestaohospital} &  33 &  3506 &  20 &  &  MongoDB \\
\emph{http-patch-spring} &  30 &  1450 &  6 &  &   \\
\emph{languagetool} &  1385 &  174781 &  2 &  &   \\
\emph{market} &  124 &  9861 &  13 & \checkmark &  H2 \\
\emph{microcks} &  471 &  66186 &  88 & \checkmark &  MongoDB \\
\emph{ocvn} &  526 &  45521 &  258 & \checkmark &  H2;MongoDB \\
\emph{ohsome-api} &  87 &  14166 &  134 &  &  OSHDB \\
\emph{pay-publicapi} &  377 &  34576 &  10 & \checkmark &  Redis \\
\emph{person-controller} &  16 &  1112 &  12 &  &  MongoDB \\
\emph{proxyprint} &  73 &  8338 &  74 & \checkmark &  H2 \\
\emph{quartz-manager} &  129 &  5068 &  11 & \checkmark &   \\
\emph{reservations-api} &  39 &  1853 &  7 & \checkmark &  MongoDB \\
\emph{rest-ncs} &  9 &  605 &  6 &  &   \\
\emph{rest-news} &  11 &  857 &  7 &  &  H2 \\
\emph{rest-scs} &  13 &  862 &  11 &  &   \\
\emph{restcountries} &  24 &  1977 &  22 &  &   \\
\emph{scout-api} &  93 &  9736 &  49 & \checkmark &  H2 \\
\emph{session-service} &  15 &  1471 &  8 &  &  MongoDB \\
\emph{spring-actuator-demo} &  5 &  117 &  2 & \checkmark &   \\
\emph{spring-batch-rest} &  65 &  3668 &  5 &  &   \\
\emph{spring-ecommerce} &  58 &  2223 &  26 & \checkmark &  MongoDB;Redis;Elasticsearch \\
\emph{spring-rest-example} &  32 &  1426 &  9 &  &  MySQL \\
\emph{swagger-petstore} &  23 &  1631 &  19 &  &   \\
\emph{tiltaksgjennomforing} &  472 &  27316 &  79 & \checkmark &  PostgreSQL \\
\emph{tracking-system} &  87 &  5947 &  67 & \checkmark &  H2 \\
\emph{user-management} &  69 &  4274 &  21 &  &  MySQL \\
\emph{webgoat} &  355 &  27638 &  204 & \checkmark &  H2 \\
\emph{youtube-mock} &  29 &  3229 &  1 &  &   \\
\midrule
Total 36 & 6465 & 657162 & 1487 & 15 & 25 \\
\bottomrule 
\end{tabular} 

}
\end{table}

To answer {\bf RQ2}, we need a selection of real-world APIs.
We selected our own WFD~\cite{sahin2025wfc}, previously known as EMB~\cite{icst2023emb}.
This is a curated selection of APIs for experimentation that has been maintained since 2017,
where each year new APIs are added (e.g., based on what used by the community in empirical studies involving REST APIs).
For this study, we used its latest version 4.3.0~\cite{zenodo430wfd} at the time of writing, which contains 36 open-source REST APIs.
These APIs have various degrees of complexity, from small simple APIs to large APIs coming from the public administration from around the world (e.g., Norway, UK, Germany and Thailand).
Table~\ref{tab:suts}  shows some statistics on these 36 APIs, including for example their number of HTTP endpoints.

As \evo is a mature tool used in several industrial contexts (for example at Fortune 500 enterprises like Volkswagen~\cite{icst2025vw} and Meituan~\cite{zhang2025fuzzing}), we could use the whole entirety of WFD.
In our empirical study, we did not need to exclude any API from WFD due any technical limitations.

We ran \evo on each of the 36 APIs of WFD for 1 hour, repeating these experiments 5 times.
In total, those experiments took at least $36 \times 5 = 180$ hours, i.e., at least 7 days if run sequentially.

To answer {\bf RQ3}, we only looked at the data from the experiments on the real-world APIs in WFD.
Computational overhead results on the \totoracles small artificial APIs developed for {\bf RQ1} would had been of limited scientific interest.

\subsection{Empirical Results}
\label{sub:}

In the experiments for {\bf RQ1}, all faults in all the \totoracles artificial APIs were found in all runs, with no exceptions.
Due to their reliability, these APIs have now been added to the regression test suites of \evo itself, as end-to-end tests~\cite{arcuri2023building}.
This means that, when any change is pushed to the codebase of \evo, its CI currently on GitHub Actions automatically runs \evo on those APIs, and verifies that all the injected faults can still be found.
If not, it means that the code change has broken this functionality in \evo.

\begin{result}
{\bf RQ1}: With our novel techniques, all the injected faults in the \totoracles artificial APIs can be reliably found.
\end{result}

\begin{table}[p]
\centering
\caption{ Detected faults on the 36 APIs from WFD.
\label{tab:results}
}
\vspace{-1.5\baselineskip}
\begin{adjustbox}{width=.95\textwidth,center}
\begin{tabular}{lrrrrrrrr }\\ 
\toprule 
 SUT & \#Endp   & F900 & F902 & F903 & F904 & F905 & F907 & F908  \\ 
\midrule 
\emph{bibliothek} & 8 &  &  &  &  &  &  &  \\ 
\rowcolor{gray!30} \emph{blogapi} & 52 &  &  &  & 0.6 &  &  & 4.0 \\ 
\emph{catwatch} & 14 &  &  &  &  &  &  &  \\ 
\rowcolor{gray!30} \emph{cwa-verification} & 5 &  &  &  &  &  &  &  \\ 
\emph{erc20-rest-service} & 13 &  &  &  &  &  &  &  \\ 
\rowcolor{gray!30} \emph{familie-ba-sak} & 183 &  &  &  &  &  &  &  \\ 
\emph{features-service} & 18 &  &  &  &  &  & 1.0 & 11.0 \\ 
\rowcolor{gray!30} \emph{genome-nexus} & 23 &  &  &  &  &  &  & 2.0 \\ 
\emph{gestaohospital} & 20 &  &  &  & 1.0 &  &  & 3.0 \\ 
\rowcolor{gray!30} \emph{http-patch-spring} & 6 &  &  &  &  &  &  &  \\ 
\emph{languagetool} & 2 &  &  &  &  &  &  & 1.0 \\ 
\rowcolor{gray!30} \emph{market} & 13 & 1.0 &  &  & 0.2 &  &  &  \\ 
\emph{microcks} & 88 &  &  &  & 1.0 &  &  & 9.0 \\ 
\rowcolor{gray!30} \emph{ocvn} & 258 &  &  &  &  &  &  &  \\ 
\emph{ohsome-api} & 134 &  &  &  &  &  &  & 67.0 \\ 
\rowcolor{gray!30} \emph{pay-publicapi} & 10 &  &  &  &  &  &  &  \\ 
\emph{person-controller} & 12 &  &  &  &  &  &  & 2.0 \\ 
\rowcolor{gray!30} \emph{proxyprint} & 74 &  &  &  &  &  &  & 6.0 \\ 
\emph{quartz-manager} & 11 &  &  &  &  &  &  &  \\ 
\rowcolor{gray!30} \emph{reservations-api} & 7 &  &  &  &  &  &  & 2.0 \\ 
\emph{rest-ncs} & 6 &  &  &  &  &  &  &  \\ 
\rowcolor{gray!30} \emph{rest-news} & 7 &  &  &  &  &  &  &  \\ 
\emph{rest-scs} & 11 &  &  &  &  &  &  &  \\ 
\rowcolor{gray!30} \emph{restcountries} & 22 &  &  &  &  &  &  &  \\ 
\emph{scout-api} & 49 & 5.4 &  &  & 2.6 &  &  &  \\ 
\rowcolor{gray!30} \emph{session-service} & 8 &  &  &  &  &  &  & 1.0 \\ 
\emph{spring-actuator-demo} & 2 &  &  &  &  &  &  &  \\ 
\rowcolor{gray!30} \emph{spring-batch-rest} & 5 &  &  &  &  &  &  &  \\ 
\emph{spring-ecommerce} & 27 &  &  &  &  &  &  &  \\ 
\rowcolor{gray!30} \emph{spring-rest-example} & 9 &  &  &  & 0.8 &  & 0.2 &  \\ 
\emph{swagger-petstore} & 19 &  &  &  &  &  &  &  \\ 
\rowcolor{gray!30} \emph{tiltaksgjennomforing} & 79 &  &  &  &  &  &  & 6.0 \\ 
\emph{tracking-system} & 67 &  &  &  & 2.4 & 4.6 &  & 17.0 \\ 
\rowcolor{gray!30} \emph{user-management} & 21 &  & 0.8 & 0.8 & 0.8 &  &  & 2.0 \\ 
\emph{webgoat} & 204 & 7.0 &  &  & 0.6 &  &  &  \\ 
\rowcolor{gray!30} \emph{youtube-mock} & 1 &  &  &  &  &  &  &  \\ 
\midrule 
Mean  & 41 & 0.4 & 0.0 & 0.0 & 0.3 & 0.1 & 0.0 & 3.7 \\ 
Median  & 14 & 0.0 & 0.0 & 0.0 & 0.0 & 0.0 & 0.0 & 0.0 \\ 
Sum  & 1488 & 13.4 & 0.8 & 0.8 & 9.9 & 4.6 & 1.2 & 133.0 \\ 
\midrule 
\#SUTs & 36 & 3 & 1 & 1 & 9 & 1 & 2 & 14 \\ 
\bottomrule 
\end{tabular} 

\end{adjustbox}
\end{table}

Table~\ref{tab:results} shows the results of the experiments on the 36 APIs of WFD, averaged out of 5 runs.
A total of \foundfaults faults were found, where the large majority is of type ``\invalidallow'' (908).
We manually verified all these \foundfaults faults, to make sure that they were indeed actual faults and not false positives, or that there was any error or missing edge-case in the implementation of our novel oracles.

All types of faults were found, but two:
``\sideeffectsfailedmodification'' (901) and ``\invalidmergepatch'' (906).
It might be that our techniques are not good enough to find these types of faults in these APIs, or simple there is no fault of such types in those APIs.
We cannot know for sure.
However, among the \totoracles defined oracles, it is not unexpected that those two types in particular find the least number of faults.

The case of ``\invalidmergepatch'' (906) is rather straightforward: the method \texttt{PATCH} was introduced much later to the specs of HTTP, and therefore its use is not as popular as the other HTTP methods.
There are only a few endpoints in WFD that use \texttt{PATCH}.
So, statistically, we would reasonably expect to find fewer faults specific to a \texttt{PATCH} compared to for example a \texttt{PUT}.

The case of ``\sideeffectsfailedmodification'' (901)  is more nuanced.
Typically, REST APIs are stateless, where any needed  state is handled outside of the API (e.g., in databases).
A typical API would validate its inputs before processing its data, and make modifications to its state.
Often, for efficiency, in SQL databases all these modifications would be committed into a single, atomic transaction.
A ``\sideeffectsfailedmodification'' fault would manifest if some input validations would be executed \emph{after} some data has already been processed and saved/modified.
This could happen for example when an API interacts with more than one external service (e.g., multiple databases), and wrongly processes them one at a time, instead of validating all inputs before doing any processing.
However, this type of scenario does not seem common/present in the APIs of WFD.

It is not feasible to discuss all the faults found in these experiments.
However, there are two in particular that are interesting to discuss in more details:
``\invalidlocation'' (907) in \emph{features-service}
and
``\nonidempotentput'' (905) in \emph{tracking-system}.

\begin{figure}
\begin{lstlisting}[language=Java,numbers=left,xleftmargin=2em]
@POST
@Path("requires")
public Response addRequiresConstraintToProduct(
                 @PathParam("productName") String productName,
                 @FormParam("sourceFeature") String sourceFeatureName,
                 @FormParam("requiredFeature") String requiredFeatureName
) throws URISyntaxException {
    FeatureConstraint newConstraint = productsService.addRequiresConstraintToProduct(productName, sourceFeatureName, requiredFeatureName);
    return Response.created(new URI("/products/" + productName + "/constraints/" + newConstraint.getId())).build(); ©\label{line:withs}©
}

@POST
@Path("excludes")
public Response addExcludesConstraintToProduct(
                 @PathParam("productName") String productName,
                 @FormParam("sourceFeature") String sourceFeatureName,
                 @FormParam("excludedFeature") String excludedFeatureName
) throws URISyntaxException {
    FeatureConstraint newConstraint = productsService.addExcludesConstraintToProduct(productName, sourceFeatureName, excludedFeatureName);
    return Response.created(new URI("/products/" + productName + "/constraint/" + newConstraint.getId())).build(); ©\label{line:nos}©
}
\end{lstlisting}
\caption{\label{fig:features}
Code extract from the class \texttt{ProductsConstraintsResource} in the API \emph{features-service}, showing the implementation of two endpoints that return a \texttt{Location} header in their responses.
}
\end{figure}

Figure~\ref{fig:features} shows the implementation for the endpoints \newline
\texttt{POST:/products/\{productName\}/constraints/excludes}
and \newline
\texttt{POST:/products/\{productName\}/constraints/requires}.
Both endpoints return a \texttt{Location} header in their response, representing the location of where the new resources are created.

Naively doing a \texttt{GET} request on the URL returned by the \texttt{requires} endpoint would be wrong, as the API has no \texttt{GET} operation for it, it rather has a \texttt{DELETE} endpoint for it.
Having only a \texttt{DELETE} but not a \texttt{GET} might be considered a rather awkward design choice, but it is technically not a problem for HTTP.
This is one ``edge-case'' that \evo can correctly handle without creating any false positive.

However, when it comes to what is returned by the \texttt{excludes} endpoint, there is nothing matching in the schema.
As such, the link is treated as an external one, and a \texttt{GET} request is sent.
As it fails, this is marked by \evo as a  ``\invalidlocation'' (907) fault.
A perceptive reader can see the difference in the missing letter ``\texttt{s}'' between the plural
\texttt{"/constraints/"} on Line~\ref{line:withs}
and the singular
\texttt{"/constraint/"} on Line~\ref{line:nos}.
We can safely claim that the missing ``\texttt{s}'' is a software fault on Line~\ref{line:nos}.

For the API \emph{tracking-system}, \evo can generate tests with the following structure:

\begin{lstlisting}
PUT  /app/api/assignments/update   -> 200
GET  /app/api/assignments          -> 200
PUT  /app/api/assignments/update   -> 200
GET  /app/api/assignments          -> 200
\end{lstlisting}

The \texttt{PUT} is implemented as adding a new element to the collection \texttt{assignments}.
As the array returned by the second \texttt{GET} has one extra element compared to what returned in the first \texttt{GET}, then \evo marks this scenario as a ``\nonidempotentput'' (905) fault.
As previously discussed in Section~\ref{sub:nonidempotentput}, this type of fault is \emph{critical}.
Not only this is a major divergence from REST design principles, i.e., having verbs such as \texttt{update} in the resource path, where a more conventional/idiomatic design to add new elements to collections would had been to have that operation in a \texttt{POST} on  \texttt{/app/api/assignments}.
But also, the major problem here is that several duplicates could be added to the collections with a single \texttt{PUT} request, at random, every few hundreds/thousands requests, without the clients being informed about it.

\begin{result}
{\bf RQ2}: A total of \foundfaults faults of 7 out of \totoracles types were found on the 36 APIs of WFD, including critical faults like ``\nonidempotentput''.
\end{result}

\begin{table}[p]
\centering
\caption{ For the 36 APIs in WFD, we report the obtained 2xx coverage of their endpoints, and the overhead in seconds of our new HTTP-semantics-checks phase.
\label{tab:overhead}
}
\vspace{-1.5\baselineskip}
\begin{adjustbox}{max totalheight=.95\textheight,center}
\begin{tabular}{ l rrr}\\ 
\toprule 
SUT & \#Endpoints & \% 2xx Coverage & Overhead \\ 
\midrule 
\emph{bibliothek} &  8 &  12.5  &  0.0  \\ 
\rowcolor{gray!30} \emph{blogapi} &  52 &  35.4  &  8.8  \\ 
\emph{catwatch} &  14 &  48.2  &  0.0  \\ 
\rowcolor{gray!30} \emph{cwa-verification} &  5 &  96.0  &  0.0  \\ 
\emph{erc20-rest-service} &  13 &  7.7  &  0.0  \\ 
\rowcolor{gray!30} \emph{familie-ba-sak} &  183 &  54.1  &  5.8  \\ 
\emph{features-service} &  18 &  100.0  &  5.8  \\ 
\rowcolor{gray!30} \emph{genome-nexus} &  23 &  69.6  &  0.0  \\ 
\emph{gestaohospital} &  20 &  31.0  &  1.6  \\ 
\rowcolor{gray!30} \emph{http-patch-spring} &  6 &  100.0  &  1.2  \\ 
\emph{languagetool} &  2 &  100.0  &  0.0  \\ 
\rowcolor{gray!30} \emph{market} &  13 &  73.8  &  3.6  \\ 
\emph{microcks} &  88 &  40.9  &  4.2  \\ 
\rowcolor{gray!30} \emph{ocvn} &  258 &  83.1  &  3.2  \\ 
\emph{ohsome-api} &  134 &  3.6  &  3.0  \\ 
\rowcolor{gray!30} \emph{pay-publicapi} &  10 &  68.0  &  0.0  \\ 
\emph{person-controller} &  12 &  43.8  &  0.0  \\ 
\rowcolor{gray!30} \emph{proxyprint} &  74 &  66.9  &  104.5  \\ 
\emph{quartz-manager} &  11 &  36.4  &  1.0  \\ 
\rowcolor{gray!30} \emph{reservations-api} &  7 &  57.1  &  0.8  \\ 
\emph{rest-ncs} &  6 &  100.0  &  0.0  \\ 
\rowcolor{gray!30} \emph{rest-news} &  7 &  85.7  &  0.2  \\ 
\emph{rest-scs} &  11 &  100.0  &  0.0  \\ 
\rowcolor{gray!30} \emph{restcountries} &  22 &  100.0  &  0.0  \\ 
\emph{scout-api} &  49 &  82.0  &  4.4  \\ 
\rowcolor{gray!30} \emph{session-service} &  8 &  65.0  &  0.2  \\ 
\emph{spring-actuator-demo} &  2 &  100.0  &  0.0  \\ 
\rowcolor{gray!30} \emph{spring-batch-rest} &  5 &  100.0  &  0.0  \\ 
\emph{spring-ecommerce} &  27 &  46.7  &  0.0  \\ 
\rowcolor{gray!30} \emph{spring-rest-example} &  9 &  80.0  &  7.8  \\ 
\emph{swagger-petstore} &  19 &  76.8  &  0.0  \\ 
\rowcolor{gray!30} \emph{tiltaksgjennomforing} &  79 &  8.9  &  1.0  \\ 
\emph{tracking-system} &  67 &  71.3  &  8.4  \\ 
\rowcolor{gray!30} \emph{user-management} &  21 &  72.6  &  1.8  \\ 
\emph{webgoat} &  204 &  82.5  &  22.2  \\ 
\rowcolor{gray!30} \emph{youtube-mock} &  1 &  40.0  &  0.0  \\ 
\midrule 
Average  &  &  65.0  &  5.3  \\ 
Median  &  &  70.5  &  0.5  \\ 
\bottomrule 
\end{tabular} 

\end{adjustbox}
\end{table}

To answer {\bf RQ3}, Table~\ref{tab:overhead} shows the computational overhead, in seconds, of our novel techniques,  averaged out of 5 runs.
As our novel techniques depend on the success of the fuzzing at creating valid 2xx requests, we also report statistics on such data.

For most cases, our techniques take just a few seconds, where the median time is only $0.5$ seconds.
There are only two cases taking more than 10 seconds, which are
22 for \emph{webgoat},
and
104 (less than 2 minutes) for \emph{proxyprint}.
No API reached the timeout of 6 minutes.

For a large API with 204 endpoints such as \emph{webgoat}, where we can successfully 2xx cover 82.5\% of them, it is not surprising that our novel techniques would need more time to evaluate many different scenarios.
However, the case of \emph{proxyprint} is more peculiar.
The main reason for the much higher overhead is that, on this API, each single HTTP call is expensive to run, and takes significantly longer to execute compared to other APIs.
Even if our novel techniques do not need to evaluate many new constructed scenarios, if those though take long time to run, the final overhead might be non-negligible.
Still, less than 2 minutes can be considered as a small amount compared to the typical budget of 1 hour for fuzzing REST APIs.

\begin{result}
{\bf RQ3}: In most cases, the computational overhead of our novel techniques is negligible, being only of a few seconds, with a median time of just $0.5$ seconds.
\end{result}

\section{Threats To Validity}
\label{sec:threats}

The \totoracles rules we defined based on the specifications of HTTP are based on our interpretation of the HTTP semantics defined in different RFC documents, like 9110.
As for any human activity, there is always a chance of misinterpretations that were not caught during peer-reviewing.

Likewise, the check for possible false positives was a manual process, which is as well potentially prone to human error and disagreements.

The results of our experiments are based on a code implementation, which could had been faulty, and that so could had led to some incorrect claims.
To mitigate such risk, such implementation has been thoroughly tested.
As it is released open-source, anyone can review how it is implemented and how it is tested.

Our results on a large set of 36 APIs from WFD increase the chance of possibly generalize to other open-source APIs.
However, there is no guarantee that our results would hold on closed-source APIs developed in industry.
In the future, feedback from practitioners in industry that use \evo (e.g., like in companies such as Volkswagen and Meituan) will be useful to validate and/or refine the oracles presented in this paper.

\section{Conclusions}
\label{sec:conclusions}

Fuzzing REST APIs is a popular topic in the scientific literature, with tens of different fuzzers that have being presented.
Typically, such fuzzers detect faults based on server errors (HTTP 500 status responses).
However, several different automated oracles could be defined to find different kinds of faults that do not necessarily lead to crashes like HTTP 500.

In this paper, we have  presented novel techniques using \totoracles automated oracles to find new types of HTTP-compliance faults that could not be found with existing techniques.
Experiments on an established dataset such as WFD, using all of its 36 APIs, show that it was possible to detect \foundfaults faulty endpoints.
These faults include detecting critical issues such as non-idempotent implementations of \texttt{PUT} endpoints.

For our experiments in this paper, our techniques have been implemented in the state-of-the-art, open-source fuzzer \evo.
However, our techniques are not tailored to \evo, and could be integrated in any other REST API fuzzer, for both \emph{black-box} and \emph{white-box} testing.

Future work will aim at designing new automated oracles to detect further faults in REST APIs, as well as improving the internal fuzzing engines (e.g., to achieve better coverage) of fuzzers such as \evo.
Follow up experience reports in industry, among practitioners that already use \evo (e.g., at Volkswagen~\cite{icst2025vw},  Meituan~\cite{zhang2025fuzzing} and several others~\cite{sahin2026using}), will help to better analyze the practical impact of our novel techniques presented in this paper.

\section*{Acknowledgments}
This work is funded by the European Research Council (ERC) under the European Union’s Horizon 2020 research and innovation programme (EAST project, grant agreement No. 864972).
Omur Sahin is supported by the TÜBİTAK 2219 International Postdoctoral Research Fellowship Program (Project ID: 1059B192300060).

\section*{Data Availability}

All the techniques presented in this paper are implemented as part of \evo.
The fuzzer \evo is open-source on GitHub,\footnote{\url{https://github.com/WebFuzzing/EvoMaster}}
where each new release is automatically published on Zenodo for long-term storage (e.g.,~\cite{zenodo611evomaster}).

WFD is open-source on GitHub,\footnote{\url{https://github.com/WebFuzzing/Dataset}}
with as well each new release automatically uploaded to Zenodo for long-term storage (e.g.,~\cite{zenodo430wfd}).


\bibliographystyle{ACM-Reference-Format} 

%

\bibliography{../../../papers}


\end{document}